\documentstyle[aps,psfig,12pt,floats,tighten]{revtex}
\input isolatin1.sty

\begin{document}

\title{Dynamic Fluctuations of Semiflexible Polymers}

\author{R.  Everaers$^{*\dagger}$, F. Jülicher$^*$, A. Ajdari$^+$, 
  A.C.  Maggs$^+$ }
\address{$^+$ Physico-Chimie Th\'eorique, 
ESPCI, 10 rue Vauquelin, 75231 Paris Cedex 05, France. \\
  $^*$ Physico-Chimie Curie, 26 rue d'Ulm, 75248 Paris Cedex 05,
  France. \\
  $^\dagger$ Max-Planck-Institut f\"ur Polymerforschung,
  Postfach 3148, D-55021 Mainz, Germany. \\
  (31 July 1998)
%  {\rm (RE's draft July 31)}
  }
\address{
%\begin{minipage}{5.55in}
\begin{abstract}\hskip 0.15in
  We develop a scaling theory to describe dynamic fluctuations of a
  semiflexible polymer and find several distinct regimes.  We
  performed simulations to characterize the longitudinal and
  transverse dynamics; using ensemble averaging for a range of
  different degrees of coarse-graining we avoid the problems of slow
  equilibration often encountered in simulations. We find that the
  longitudinal fluctuations of a semiflexible object
  scales as $t^{7/8}$. These fluctuations are correlated over a length
  which varies as $t^{1/8}$.  Our results are pertinent to the
  interpretation of high frequency microrheology experiments in actin
  solutions.
\end{abstract}
%\end{minipage}
\vspace*{-0.5cm} }
\maketitle
Experiments on actin filaments have awakened interest in the
dynamics of semiflexible polymers.  Although the collective behavior
in the semidilute regime has attracted much attention \cite{sackmann},
the single filament problem displays surprisingly rich dynamic
features absent in the case of ``flexible'' polymers.  This is a
consequence of the highly anisotropic nature of the semiflexible
polymer.  The static anisotropy is characterized by the exponent for
the growth of transverse fluctuations, $r_{\perp}$ as a function of
the filament length, $L<\kappa$ \cite{odjik}, $r_{\perp}^2 \sim
L^3/\kappa$ where $\kappa$ is the persistence length of the filament.
The dynamics is commonly described by a Langevin equation for the
transverse fluctuations \cite{note},
\begin{equation}
{\partial r_{\perp} \over \partial t} = -\kappa { \partial^4 r_{\perp} \over
\partial s^4} + f_\perp(s,t)
\label {langevin}
\end{equation}
Here, $f_\perp$ denotes a transverse stochastic force per unit length
applied to the filament, $s$ is a curvilinear coordinate and
hydrodynamic interactions are neglected on the ground that they only
induce logarithmic corrections.  After time $t$, the filament is
equilibrated over a length $l_1(t)\sim (\kappa t)^{1/4}$.  This
results in a scaling of the transverse dynamics first measured in
dynamic light scattering \cite{schmidtthesis}, where it is seen that
the transverse motion of a monomer on a filament varies as
\cite{farge}
\begin{equation}
\langle \delta r_{\perp}(t)^2\rangle \sim l_1(t)^3/\kappa 
\sim t^{3/4}/\kappa^{1/4}
\label {transverse}
\end{equation}
The longitudinal fluctuations of a filament are even more subtle.  We
need to impose strict incompressibility on the filament, which is
violated in the simplest linearized equations such as
(\ref{langevin}).  This is a long standing and delicate problem in the
description of worm like polymers \cite{harris}.  Recently scaling in
$t^{3/4}$ has been predicted \cite{fred,morse} for the longitudinal
fluctuations of a filament, with however an amplitude smaller by a
factor $L/\kappa$ than in (\ref{transverse}); 
we can also find this result with a simple scaling
argument: Due to the local conservation of filament length the
transverse fluctuations of a segment of length $l$ result in parallel
fluctuations of $\delta r_{l}\simeq {1\over 2}\int_0^l d s (\partial
r_{\perp}/ds)^2 $ comparable to $\langle \delta r_{l} ^2\rangle \sim
l^4/\kappa^2$. Each section of length $l_1(t)$ of the filament is
independent so that we can add the fluctuations of $m=L/l_1(t)$
segments giving the motion of the end
\begin{equation}
\langle \delta r_{\parallel}(t)^2\rangle  \sim m \,\delta r_{l}^2 =
(L/l_1) (l_1^4/\kappa^2)
= L t^{3/4}/\kappa^{5/4}
\label {long1}
\end{equation}

While applicable to the case of a filament in a macroscopic shear
field \cite{morse} this result is problematic for local probes of the
dynamics. In particular the divergence of the fluctuations with the
filament length must surely break down at some point.  A similar
problem exists in the athermal dynamics of semiflexible filaments
\cite{nelson,ajdari1} where the longitudinal friction of the filament
modifies the dynamics.  It is the purpose of this letter to explore
the importance of longitudinal dissipation and the incompressibility
constraint in the dynamics of the fully thermalised chain.

In the derivation leading to (\ref{long1}), longitudinal friction is
absent and all the different segments of length $l_1(t)$ contribute by
their longitudinal fluctuations to the motion of the end.  However,
the shortening (or extension) of a segment also requires the
longitudinal motion of its neighbors.  As a consequence, longitudinal
friction limits the number of segments which can contribute within a
finite time.  Typically this limitation should not be significant if
the longitudinal diffusion of the whole polymer (diffusing as $r^2 =
t/L$) is so fast that there is no hindrance to the cumulation of
fluctuations.  This introduces a criterion for the validity of Eq.
(\ref{long1}):
\begin{math}
  L t^{3/4}/ \kappa^{5/4} < t /L
\label{ineq}
\end{math}
or $L < l_2(t)= t^{1/8}\kappa^{5/8}$.  For longer filaments only a
section of length $l_2(t)$ can contribute to the fluctuation of the
end point. This limitation thus amounts in substituting $l_2$ for $L$
in (\ref{long1}):
\begin{equation}
\langle \delta r_{\parallel}(t)^2\rangle \sim t^{7/8}/\kappa^{5/8}
\label{long2}
\end{equation}

A similar result can be found from an analysis of the linear
response of a filament to a weak constant longitudinal force $f_{\|}$
using the fluctuation dissipation theorem \cite{forster}.  Equation
(\ref{long1}) predicts that in the presence of a force the end drifts
as $\delta r_{\|}(t)\sim f_{\|} L t^{3/4}/\kappa^{5/4}$.  However we
cannot set the whole filament into motion at once: the typical
velocity scales as $v_{\|} \sim \delta r_{\|}/t \sim f_{\|}
L/\kappa^{5/4}t^{1/4}$, and the total drag $Lv_{\|}$ can not be larger
than the applied force. This requirement again reads $L<l_2(t)$.  For
a long filament the tension propagates a distance $l_2(t)$.  The
corresponding section along which tension has propagated is set into
motion at a velocity of the order of $v_{\|}\approx f_{\|}/l_2$ and
the end drifts a distance $\delta r_{\|}(t)\sim f_{\|}
t^{7/8}/\kappa^{5/8}$, in agreement with (\ref{long2}).

Thus a surprising conclusion is that there are {\em two dynamic length
  scales}\/ which play important roles in the dynamics of filaments. Our
main result is thus that the dynamics obey a pair of scaling relations
with  different scaling arguments.
In particular, the fluctuations  of the end of a filament
are expected to follow: 
\begin{eqnarray}
\langle \delta r_{\parallel} (t)^2 \rangle &=&  {t^{7/8} \over
\kappa^{5/8} }{\mathcal Q}\left({t^{1/8} \kappa^{5/8} \over L}\right) 
\label{scale} \\ 
 \langle  \delta r_{\perp} (t)^2 \rangle &=&
  {t^{3/4} \over \kappa^{1/4} }{\mathcal W}\left({t^{1/4} 
\kappa^{1/4}\over L}\right)
\label{scale2}
\end{eqnarray}
%We emphasize the unusual feature of having two different scaling
%arguments in (\ref{scale}, \ref{scale2}).  
In the rest of this paper
we present computer simulations of a model filament in order to check
these scaling arguments.
% and determine the function $\mathcal{Q}$.

We have performed simulations of a semiflexible polymer in two dimensions
imposing a constraint on the contour length using the technique
described in \cite{hinch}.  The polymer is discretized with sequence
of beads with positions $\vec r_i$, $i\in 0\ldots n$, with fixed
distance $b=|\vec r_i-\vec r_{i-1}|$ and normalized bond vectors $\vec
d_i = (\vec r_i - \vec r_{i-1})/b$.  The angles $\theta_i$
characterizing the bond directions are coupled by simple angular
springs:
\begin{equation}
E = \frac{1}{2} \sum_{i=1}^{n-1} \frac{\kappa}{b} (\theta_i -\theta_{i+1})^2
\label{energy}
\end{equation}
The beads move against an isotropic friction $-b \partial{\vec
  r}_i/\partial t$ under the influence of the forces due to the
angular springs, $- \partial E/\partial \vec r_i$ and stochastic
forces $\vec F^{\rm ra}_i$:
\begin{equation}
b \frac{\partial{\vec r}_i}{\partial t} = - \frac{\partial E}{\partial\vec r_i} + \vec F^{\rm ra}_i
+ T_{i+1} \vec d_{i+1} -  T_i \vec d_i 
\label{equa}
\end{equation}
The tensions $T_i$ play the role of Lagrange multipliers whose values
are calculated at each time step from the condition that the bond
lengths are equal to $b$.

The shortest characteristic time of the model is approximately
$\tau(b) = b^4/\kappa$, while the relaxation time of a chain of
$n=L/b$ segments varies as $\tau_n \sim n^4 \tau(b)$.  The total
simulation time needed to equilibrate a chain is proportional to
$n^5$. The equilibration of long chains becomes quickly impossible;
previous studies have been limited to chains which are too short to
study detailed features of the dynamics.

We get around the problem of generating independent configurations by
simulating long chains for a time {\it far}\/ shorter than the
equilibration time but then performing ensemble averages over many
short runs. These simulations are useful because we can easily prepare
{\it fully equilibrated}\/ initial conformations for the energy eq.
(\ref{energy}) by drawing bond angles $\delta \theta =
\theta_i-\theta_{i-1}$ randomly from a Gaussian distribution
$P(\delta\theta)\sim\exp\left(-\frac\kappa{2 b}\delta\theta^2\right)$.
The choice of the segmentation then determines a { window of
  accessible time scales}. The elementary time step of the integrater
is $10^{-2} \tau(b)$; times shorter than $\tau(b)$ are affected by
discretization errors.  We generated data for times up to
$10^3\tau(b)$ with a computational effort proportional to only $n^1$.
We chose a sequence of segmentations $b_j= 2^{-j}\kappa$ and studied
chains of length $\kappa/8, \kappa/4,\kappa/2,\kappa$ with the number
of segments varying between $n=8$ and $n=1024$.  The overlap between
adjacent time windows ($\tau_{n} = 2^4 \tau_{n/2}$) provides a
convenient check on the coarse graining procedure and the scaling of
the parameters.
\begin{figure}
  \centerline{
\psfig{figure=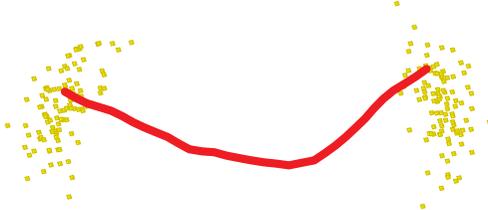,width=6.5cm}
%\pdfimage height 8cm width 8cm Fig1.png
}
\caption{Cloud of endpoints generated by simulating $N=100$
  realizations of the dynamics starting from the illustrated initial
  condition. The moments of this cloud are used to extract the
  longitudinal and transverse dynamical. We see that transverse
  fluctuations are larger in amplitude than the longitudinal
  fluctuations and are thus characterized by a larger moment.}
\end{figure}

We note that imposing a constraint on the bond lengths provides access
to longer simulation times than simulations done with stiff
longitudinal springs; we do not need to simulate the corresponding
fast but uninteresting modes which limit the integration step in any
explicit integration scheme. There are nevertheless complications: An
extended discussion can be found in \cite{hinch}. In particular, the
act of passing from elastic springs to incompressible bonds changes
the configuration space for the problem from Euclidean to Riemannian
and the distribution function changes from $\exp(-E)$ to
$\sqrt{\Delta_n} \exp(-E)$. $ \Delta_n$ is the determinant of the
Jacobian describing the transformation from Cartesian to bond angle
coordinates which can be calculated from a transfer matrix
\begin{displaymath}
\left(\matrix{
 \Delta_{i+1} \cr
\Delta_i
}\right)
=
\left(\matrix{
2 & -\cos^2 (\theta_i- \theta_{i+1}) \cr
1 &0
}\right)
\left(\matrix{
\Delta_{i} \cr
\Delta_{i-1}
}\right)
\end{displaymath}
starting from the initial vector $ (2,1)$.  To recover the
distribution function $\exp(-E)$ for stiff springs from a simulation
of a rigid rod model, it is necessary to add a pseudo
potential~\cite{fixman}, $-\frac12\log\left(\Delta_n\right)$ to $E$ 
and include the corresponding forces in (\ref{equa}).  In the
simulations described in this paper a proper calculation of these
forces is essential to ensure that we start our short runs from
initial conformations which are properly equilibrated.

\begin{figure}
  \centerline{
\psfig{figure=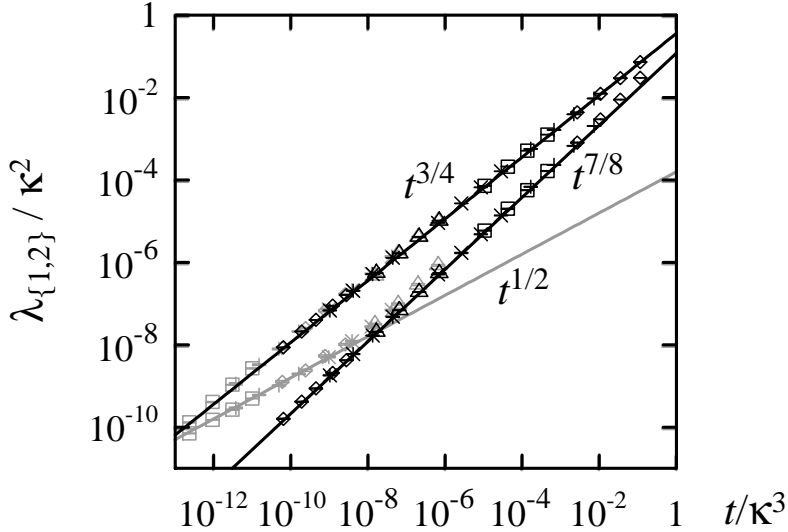,width=10.5cm}
%\pdfimage height 10cm width 10cm Fig2.png
}
\caption{Amplitude of fluctuation as a function of
  time for an incompressible filament (black symbols, $L=\kappa$) and
  a compressible filament (grey symbols, $L=\kappa/4$, modulus
  $K=10^{7}\kappa^{-1}$) determined from the moments $\lambda_1(t)$, 
  $\lambda_2(t)$ of the 2D clouds. The
  longitudinal mode scales as 
  $\lambda_2(t)=\langle \delta r_{\|}^2 \rangle \sim t^{7/8}$, 
  the transverse modes obey 
  $\lambda_1(t)=\langle \delta  r_{\perp}^2\rangle \sim t^{3/4}$. 
  For a compressible filament, a
  Rouse-like scaling 
  $\lambda_2(t)=\langle \delta r_{\|}^2\rangle \sim t^{1/2}$ is
  found for very short times. The different symbols correspond to
  different levels of coarse-graining.  }
\end{figure}
The objective of our simulations is the characterization of the
transverse and longitudinal motion of the chain ends.  For this
purpose we perform $N$ simulations (typically $N=1000$) starting from
an identical pre-equilibrated conformation; each simulation uses an
independent series of random forces.  We then record the $N$
coordinates of one end of the chain as a function of time which form
an evolving two dimensional cloud in the $(x,y)$ plane.  The cloud of
points, Fig. 1, can be characterized by two important invariants, its
two moments of inertia  $\lambda_1(t)$ and 
$\lambda_2(t)$ ($\lambda_1>\lambda_2$).  
The evolution of the $\lambda_i(t)$
characterizes the transverse and longitudinal movement of the end of
the chain (indeed $\lambda_i \sim \delta r^2$).  
We prepare a total of $M$ random realizations (typically
$M=100$) of the initial chain over which we can calculate average
properties of the cloud performing a total of $MN=10^5$ simulations, Fig 2.
For short times the evolution of the cloud is very anisotropic, and
the transverse dynamics corresponds to the larger moment 
$\lambda_1(t)$ which scales
according to eq.  (\ref{transverse}).  The smaller moment 
$\lambda_2(t)$ characterizes
the parallel motions of the filament and for short times varies in
agreement with eq.  (\ref{long2}).  For long times with $l_2>L$, a
crossover to free diffusion of the whole filament occurs.

In order to  understand better the relative importance of internal modes
and center of mass diffusion in contributing to Eq. (\ref{scale}) we
examined the joint motion of the two end-points of a chain. In a
series of $M$ simulations one generates a distribution of points in
four dimensions (4D): $\{(x_1,x_2,y_1,y_2)\}$.  The cumulants of this
distribution can be used via the fluctuation dissipation theorem to
calculate the response of the chain ends to arbitrary combinations of
end forces \cite{forster}. We are interested in the evolution of the
four moments $\{\Lambda_1>\Lambda_2>\Lambda_3>\Lambda_4\}$ of the 4D
cloud which characterize the dynamics of the whole chain.
\begin{figure}
  \centerline{
\psfig{figure=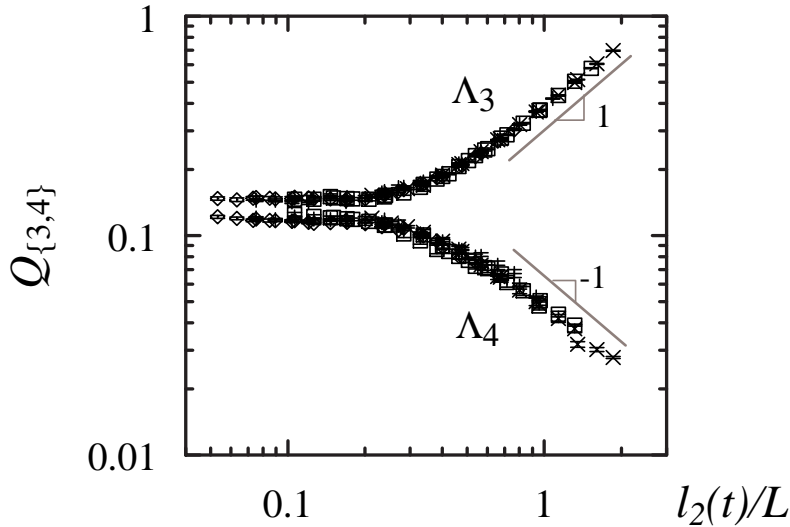,width=10.5cm}
%\pdfimage height 10cm width 10cm Fig3.png
}
\caption{Scaling behavior of the two smallest moments $\Lambda_{3,4}$ 
of the 4D clouds.
  Plotted is ${\mathcal Q}_{\{3,4\}}=\Lambda_{\{3,4\}} \kappa^{5/8}t^{-7/8}$
  as a function of the tension propagation length
  $l_2(t)=t^{1/8}\kappa^{5/8}$ for $L=\kappa$ ($\Diamond$), $\kappa/2$
  ($+$), $\kappa/4$ ($\Box$), $\kappa/8$ ($\times$).  
The plot demonstrates that the longitudinal fluctuations follow
the scaling form of Eq. (\ref{scale}).}
\end{figure}

Our picture of the propagation of tension fluctuations as introduced
above suggests the following scenario: As long as $l_2<L$, the
movement of the ends are uncorrelated. The 4D distribution factorises
and reduces to a product of the 2D case discussed above, \cite{fn1}:
the two (degenerate) smaller moments $\Lambda_3=\Lambda_4 \simeq
\lambda_2$ scale as $t^{7/8}/ \kappa^{5/8}$ and the two larger moments
$\Lambda_1=\Lambda_2 \simeq \lambda_1$ scale as
$t^{3/4}/\kappa^{1/4}$.  For longer times $l_2(t)>L$ the two ends see
each other as tension propagates along the filament.  This lifts the
degeneracy between the two smaller moments, with $\Lambda_4$ now
characterizing the end to end fluctuations so that according to (3)
$\Lambda_4 \sim L t^{3/4}/\kappa^{5/4} $, and $\Lambda_3$
characterizing the longitudinal free diffusion of the chain,
$\Lambda_3 \sim t/L$.  Figure 3 shows the moments $\Lambda_3$ and
$\Lambda_4$ plotted normalized by $t^{7/8}/\kappa^{5/8}$ as functions
of $t^{1/8}\kappa^{5/8}/L$, so that they clearly follow the scaling
form Eq. (5).

Real filaments have a longitudinal compressibility.  In order to study
the effect and to compare different methods of modeling the dynamics
we have repeated some of our simulations using harmonic potentials
$E_b=\sum_{i=1}^n K(|\vec r_i-\vec r_{i-1}|-b)^2/2b$ representing
elastic bonds with a large but finite modulus $K$.  These
simulations require an elementary time step of
$10^{-2} \frac\kappa{K b^2}\ \tau(b)$
are performed without the pseudo potential.  The internal
longitudinal dynamics are over-damped and Rouse-like ($\partial_t
r_{\|} \simeq K \partial^2_s r_{\|}$), giving fluctuations which scale
as $\delta r_{\parallel}^2\sim(t/K)^{1/2}$; the tension is correlated
over a distance $l_3(t) \sim (t K)^{1/2}$. Compressibility thus
introduces a third dynamic length $l_3$ in competition with those
already presented.  For very short times $t< \kappa^{5/3}/K^{4/3}$ the
compression modes dominate over the longitudinal modes discussed in
the first part of the article as shown in Fig 2. For longer times a
cross-over to the dynamic behavior of the constrained chain occurs.
If we assume actin filaments to be uniform rods of Young's modulus $E$
and radius $a$, $\kappa \sim E a^4$ \cite{howard} while $K\sim E a^2$.
We expect the crossover to occur for $t\sim a^{8/3}\kappa^{1/3}$ and
at a length $(a/\kappa)^{1/3}\kappa$.  For actin we estimate $t^{-1}
\simeq 1$ MHz, indicating that on time scales larger than $10^{-6}$s
the exponent $7/8$ should be observable for filaments longer than $0.1
\kappa$.

We have presented results in two dimensions where bending and torsion
are decoupled. In three dimensions we expect the coupling between
these modes produces an even richer behavior \cite{wiggins}.  In the
case of actin, a filament of length $L=\kappa$ has a torsional mode
with relaxation time $\simeq 1ms$ \cite{zimm} again relaxing with a
diffusive, Rouse like mode together with a new competing length scale.
This introduces more features which we consider in a future
publication.

Can longitudinal fluctuations of semiflexible filaments be studied
experimentally?  Following the motion of the end of a single filament
in optical microscopy may be feasible but could prove somewhat tricky.
Some form of scattering experiments sensitive to the longitudinal
motions would be preferable.  Normal dynamic light scattering
measurements are not sensitive to longitudinal motions, however one
might have some hope of introducing optical inhomogeneity \cite{paul}
in actin filaments by marking with a low concentration of dye.
Alternatively one might hope to immuno-attach nanometric gold beads to
an actin filament and follow their movement with either video or light
scattering techniques.

Several of experimental groups have studied the high frequency
fluctuations \cite{schmidt,wirtz,weitz,amblard} of actin solutions
with beads between $0.5 \mu$m and $5 \mu$m in radius. The results have
been interpreted using a theory for the macroscopic modulus of a
solution of semiflexible objects starting from eq. (\ref{long1}). We
see here that no physical cut off is needed to explain the data,
longitudinal friction leads to a dynamic length scale $l_2(t)$ which
cuts off the divergence with $L$.  The beads in these experiments are
sufficiently small that they should sample the point response of the
chain rather than being sources of uniform shear. We have seen that
the response function of a single chain has a rich scaling behavior;
any experiment will sample a mixture of $\delta r^2 \sim$ $t^1$,
$t^{3/4}$, $t^{1/2}$ and $t^{7/8}$ due to the averaging over filament
lengths and modes.  Indeed the data when examined carefully show some
curvature when plotted on a logarithmic scale. However, for
dense solutions multi-chain hydrodynamic interactions are surely
important, we do not have a detailed understanding of this regime.

Eventually tension propagation may be an issue in the understanding of
experiments analyzing motor-filament systems, as it may e.g.\ control
the response of a filament to the ``kicks'' of a collection of motors.
In particular the propagation of the tension in $t^{1/8}$ will lead to
an effective low frequency filter due to propagation delays.

The data in this article correspond to approximately $3 \times 10^6$
independent simulations each lasting an average of 2 seconds, using 3
months of CPU time on a UltraSparc workstation.  We would like to
thank Paul Chaikin, Fred Gittes, Fred Mackintosh and Jacques Prost 
for discussions on this work.


\begin{thebibliography}{99}
\bibitem{note} We use units for which $k_BT=1$ as well as $\eta = 1$,
  where $\eta$ is the friction coefficient of the filament per unit
  length.  In these units time is a volume. For actin, $1$ second
  $\approx 1 \mu$m$^3$.
\bibitem{eigen} We can estimate the range of the fictive forces
  generated by the Jacobian by calculating the two eigenvalues
  $\mu_1, \mu_2$ of the  averaged matrix.  This range
  is expected to be comparable to $1/\log(\mu_2/\mu_1)$.  When the
  polymer is flexible the interaction range is almost nearest
  neighbor generating an entropic comparable to $k_B
  T$. For semiflexible filaments $\mu_1/ \mu_2 \rightarrow 1$
  and the interaction becomes non-local with range $\sqrt{\kappa b}$.
\bibitem{fn1}  The short time difference between $\Lambda_3$ and $\Lambda_4$
in Fig. 3 is due to fluctuations of order of $1/N^{1/2}$.
\end{thebibliography}
\end{document}